\documentclass[12pt]{article}
\usepackage{latexsym}
\usepackage{epsfig}

\usepackage{graphicx}
\usepackage{epstopdf}

\usepackage{epsfig,amssymb,amsmath,amscd}

\newcommand{\bq}{\begin{equation}}
\newcommand{\eq}{\end{equation}}
\newcommand{\bqa}{\begin{eqnarray}}
\newcommand{\eqa}{\end{eqnarray}}
\newcommand{\ben}{\begin{enumerate}}
\newcommand{\een}{\end{enumerate}}
\newcommand{\bc}{\begin{center}}
\newcommand{\ec}{\end{center}}
\newcommand{\bqb}{\begin{eqnarray*}}
\newcommand{\eqb}{\end{eqnarray*}}

\def\pr#1#2#3{Phys. Rev. ${\bf{#1}}$, #2 (#3)}

\def\pl#1#2#3{Phys. Lett. ${\bf{#1}}$, #2 (#3)}

\def\np#1#2#3{Nucl. Phys. ${\bf{#1}}$, #2 (#3)}

\def\jmp#1#2#3{J. Mod. Phys. ${\bf{#1}}$, #2 (#3)}

\begin{document}
\pagenumbering{arabic}
\thispagestyle{empty}
\def\thefootnote{\fnsymbol{footnote}}
\setcounter{footnote}{1}

\vspace*{2cm}
\begin{flushright}
May 16, 2018\\
 \end{flushright}
\vspace*{1cm}

\begin{center}
{\Large {\bf Z polarization in $\gamma\gamma$,gluon-gluon $\to t\bar t Z$ 
for testing the top quark mass structure and the presence of final interactions.}}\\
 \vspace{1cm}
{\large F.M. Renard}\\
\vspace{0.2cm}
Laboratoire Univers et Particules de Montpellier,
UMR 5299\\
Universit\'{e} de Montpellier, Place Eug\`{e}ne Bataillon CC072\\
 F-34095 Montpellier Cedex 5, France.\\
\end{center}

\vspace*{1.cm}
\begin{center}
{\bf Abstract}
\end{center}

We show what type of effects on the rate of $Z_L$ production in
the  $\gamma\gamma\to t\bar t Z$ and $gluon~gluon\to t\bar t  Z$
processes could be produced by an effective scale dependent top mass 
or by final state interactions between massive particles for example
generated by the presence of dark matter.

\vspace{0.5cm}

\def\thefootnote{\arabic{footnote}}
\setcounter{footnote}{0}
\clearpage

\section{INTRODUCTION}

The motivation of this paper is to pursue what has been proposed
with the $e^+e^-\to t\bar t Z$ process \cite{eettZ}, i.e. to see the
sensitivity of the Z polarization to the origin of the top quark mass
in the  $\gamma\gamma\to t\bar t Z$ and $gluon~gluon\to t\bar t  Z$
processes.\\
As expected from the Goldstone equivalence \cite{equiv} and the $Gt\bar t$ coupling
proportional to the top quark mass, the $Z_L$ rate should reflect
any modification of this sector, for example due to top quark compositeness
or to the presence of dark matter (DM) especially connected to heavy
particles; for review about DM see
for example \cite{revDM}.\\
We will follow the same procedure as in the previous analysis.
First, in the SM case, we check that the $t\bar t  Z_L$ production is, 
at high energies (up to $m^2_Z/s$ corrections), numerically equivalent
to $t\bar t  G^0$. Incidently we also check that in the $b\bar b  Z$ case,
the $Z_L$ proportion is very small.\\
We then illustrate two possible sources of departures from the SM in
the spirit of the motivation mentioned above:\\
--- a scale dependent top mass, described by some $m_t(s)$ function,
affecting directly the $Z_L$ rate,\\
--- special additional final state interactions between heavy particles
for example originating from the environment of DM;
in fact only $t Z_L$ and  $\bar t Z_L$ pairs can influence the $Z_L$ proportion,
as a final $t\bar t$ interaction would similarly affect $Z_L$ and $Z_T$.\\

Contents: The SM description of both $\gamma\gamma\to t\bar t Z$
and $gluon~gluon\to t\bar t  Z$ is recalled in Sec.2.  The scale dependent top mass effects are illustrated in Sect.3 and the ones of final state interactions in Sect.4.
A summary is given in Section 5.\\

\section{SM description of $\gamma\gamma\to t\bar t Z$
and $gluon~gluon\to t\bar t  Z$ processes}

At Born level these processes are respectively described by the 3 and 5 diagrams
of Fig.1. The final $Z$ can have both transverse and longitudinal polarizations.
We can check the Goldstone equivalence by computing the corresponding $t\bar t G^0$ production processes replacing in each diagram the $Z$ line by a $G^0$ one.\\

Indeed the total (not term by term) $Z_L$ amplitude agrees with the $G^0$ one 
in the $p_Z>>m_Z$ limit. 
Hence the $ttG^0$ coupling
\bq
 c^L_{G^0tt}=-c^R_{G^0tt}=-i {em_t\over2s_Wm_W}
\eq
explains the direct dependence on $m_t$ of the $Z_L$ amplitude.\\
 
We have then computed the $Z_L$ ratio

\bq
R_L={\sigma(t\bar t Z_L)\over \sigma(t\bar t Z_T)+\sigma(t\bar t Z_L)}
\eq
which will constitute the basis of our study for the search of non standard effects.\\

We can illustrate the numerical agreement with the $G^0$ ratio

\bq
R_L(G^0)={\sigma(t\bar t G^0)\over \sigma(t\bar t Z_T)+\sigma(t\bar t G^0)}
\eq

In Fig.2 and 5, for the $\gamma\gamma$ and the gluon~gluon cases one can
see the accuracy of this agreement. We have done this numerical analysis 
for two angular values $\theta_Z={\pi\over6},{\pi\over2}$.\\

This direct sensitivity of the $Z_L$ amplitude to the top quark mass
can be confirmed by comparison with $b\bar b Z$ production. In this case
with the smallness of the bottom mass the rate of $Z_L$ production
is accordingly small as one can see in Fig.2 and 5.\\

So it is clear that the  $\gamma\gamma\to t\bar t Z$
and $gluon~gluon\to t\bar t  Z$ processes may be as adequate as the 
$e^+e^-\to t\bar t Z$ one (\cite{eettZ}) for studying the structure
of the top quark mass. We will now illustrate two types of non standard
effects.\\

\section{Scale dependent top mass effect}

The presence of a scale dependent top mass may arise, like
in the hadronic case and QCD, from some compositeness and its
binding interaction. 
For such compositeness and the precise cases of top quark and Higgs boson,
see for example refs.\cite{comp, Hcomp2,Hcomp3,Hcomp4,partialcomp}.\\
The possibility of such a scale dependent top mass has been mentioned in \cite{trcomp,CSMrev}.\\
In order to show its effect on the $Z$ polarization in the
$\gamma\gamma\to t\bar t Z$
and $gluon~gluon\to t\bar t  Z$ processes we will systematically replace, in the
analytic expressions of the amplitudes, the basic top mass by a
unique effective mass $m_t(s)$ expression
\bq
m_t(s)=m_t{(m^2_{th}+m^2_0)\over (s+m^2_0)}
\eq
where s is the total $\gamma\gamma$ or gluon-gluon energy squared
and $m^2_{th}$  the threshold value.\\
This is the simplest choice used in the illustrations shown in
Fig.3 and 6.
Different effective masses $m_t(x)$ depending on each subenergy
$x=s_{Zt}, s_{Z\bar t}, s_{t\bar t}$ may appear for each diagram.
This will depend on the specific compositeness model.
But our present aim is only to show what type of effect on the $Z$
polarization would be generated by such effective masses.\\
In Fig.3 and 6 we can see the reduction of $R_L$ generated by the use of
$m_t(s)$ with $m_0=2$ or $4$ TeV.\\
Indeed the effect is clear and similar in both $\gamma\gamma$ or gluon-gluon
processes.\\

\section{Dark matter final state interaction}

As already presented in (\cite{eettZ}) we will now consider final state interactions
between heavy particles.
They may also arise from the DM environment \cite{DMmass} as a subsequent consequence
of the creation of the masses, \cite{DMexch}.\\
We will follow the same phenomenology as for the  $e^+e^-\to t\bar t Z$ process.
Such final state interactions may appear between ($Zt$), ($Z\bar t$) and
($t\bar t$).\\
As discussed in  \cite{DMexch} if they are related to mass generation  they could be specific of the longitudinal gauge bosons (and correspondingly
of the Goldstone bosons).\\ 
In the present processes the ratio $R_L$ would be modified by final state processes
$Z_Lt\to Z_Lt$, $Z_L\bar t\to Z_L\bar t$ but not by the $t\bar t\to t\bar t$
one (the identification of this last interaction could be done by measurement of the 
top quark polarization in $e^+e^-\to t\bar t$ production processes as discussed in \cite{DMexch}).\\
We will now make illustrations first by simply modifying the $Z_Lt\bar t$
amplitudes by the $(1+C(s_{Zt})) (1+C(s_{Z\bar t}))$ "test factor"
with
\bq
C(x)=1+{m^2_{t}\over m^2_0}~ln{-x\over (m_Z+m_t)^2} ~~, \label{Fs}
\eq
\noindent
with the subenergies $x=s_{Zt}$ or $s_{Z\bar t}$ and $m_0=0.5$ TeV, like in \cite{DMexch}.\\
The results can be seen in Fig.4 and 7 for  $\gamma\gamma$ and gluon-gluon
processes with the curves (DMZ) compared to the standard
SM ones.\\
One will also add the possible contribution of the production
$\gamma\gamma~or~gg \to t\bar t G^0$ followed by final $G^0t\to Z_Lt$ and 
$G^0\bar t\to Z_L\bar t$ interactions. This increases the effects as shown
by curves (DMZG) in Fig.4 and 7.\\

The numerical values have no real meaning; these figures just show what
type of effect on the $Z$ polarization one could search when looking 
for the presence of dark matter.\\

\section{CONCLUSION}

In this paper we have applied to the $\gamma\gamma\to t\bar t Z$
and $gluon~gluon\to t\bar t  Z$ processes the same study as the one
previously done for the $e^+e^-\to t\bar t Z$ process.\\
The aim is to check the sensitivity of the $Z$ polarization to the
top quark mass in order to get signals of new physics related to
its origin. We have in mind top quark compositeness or the generation
of the top quark mass by the DM environment.\\
We have illustrated two types of such effects;
first the direct proportionality of the $Z_L$ ratio with
the top quark mass allowing to immediately see the occurence of a 
scale dependent effective top mass;
secondly the presence of final state interactions between heavy particles
due to the dark matter environment.\\
Illustrations with some arbitrary choices of parameters show that 
visible effects directly appear on the ratio of $Z_L$ ratio in
both $\gamma\gamma$ and gluon-gluon processes when either
of these effects occur with its specific kinematical properties.\\
Of course quantitative predictions would require a precise mass generation model\\
and the application to photon-photon and hadronic collisions will need to take into account all the detection caracteristics, see for example \cite{gammagamma, Contino, Richard}.

\newpage

\begin{figure}[p]
\vspace{-2cm}
\[
\hspace{-2cm}\epsfig{file=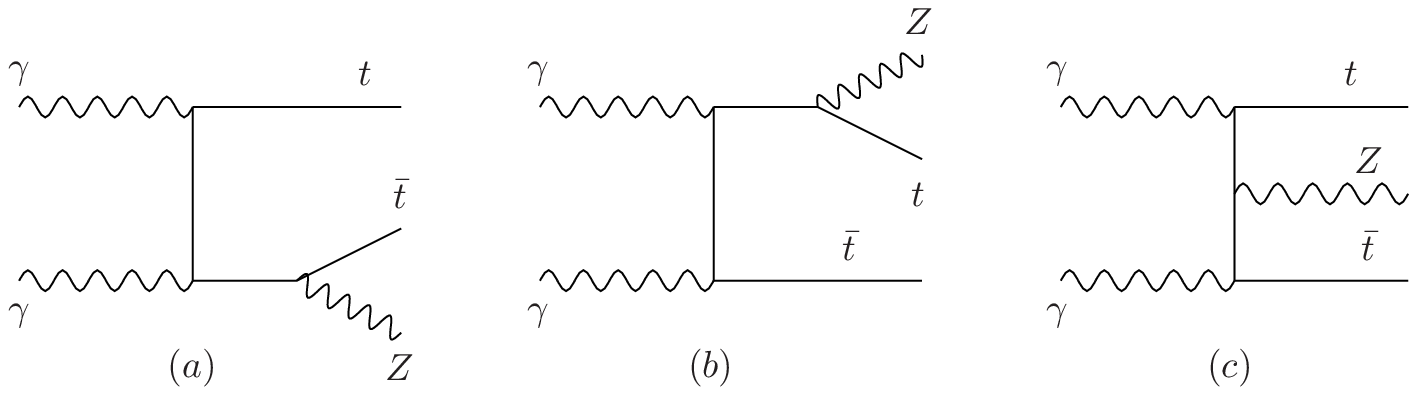 , height=4.cm}
\]\\
\vspace{-1.cm}
\hspace{4cm} SM diagrams for $\gamma\gamma \to t\bar t Z$ process\\

\vspace{1cm}
\[
\hspace{-2cm}\epsfig{file=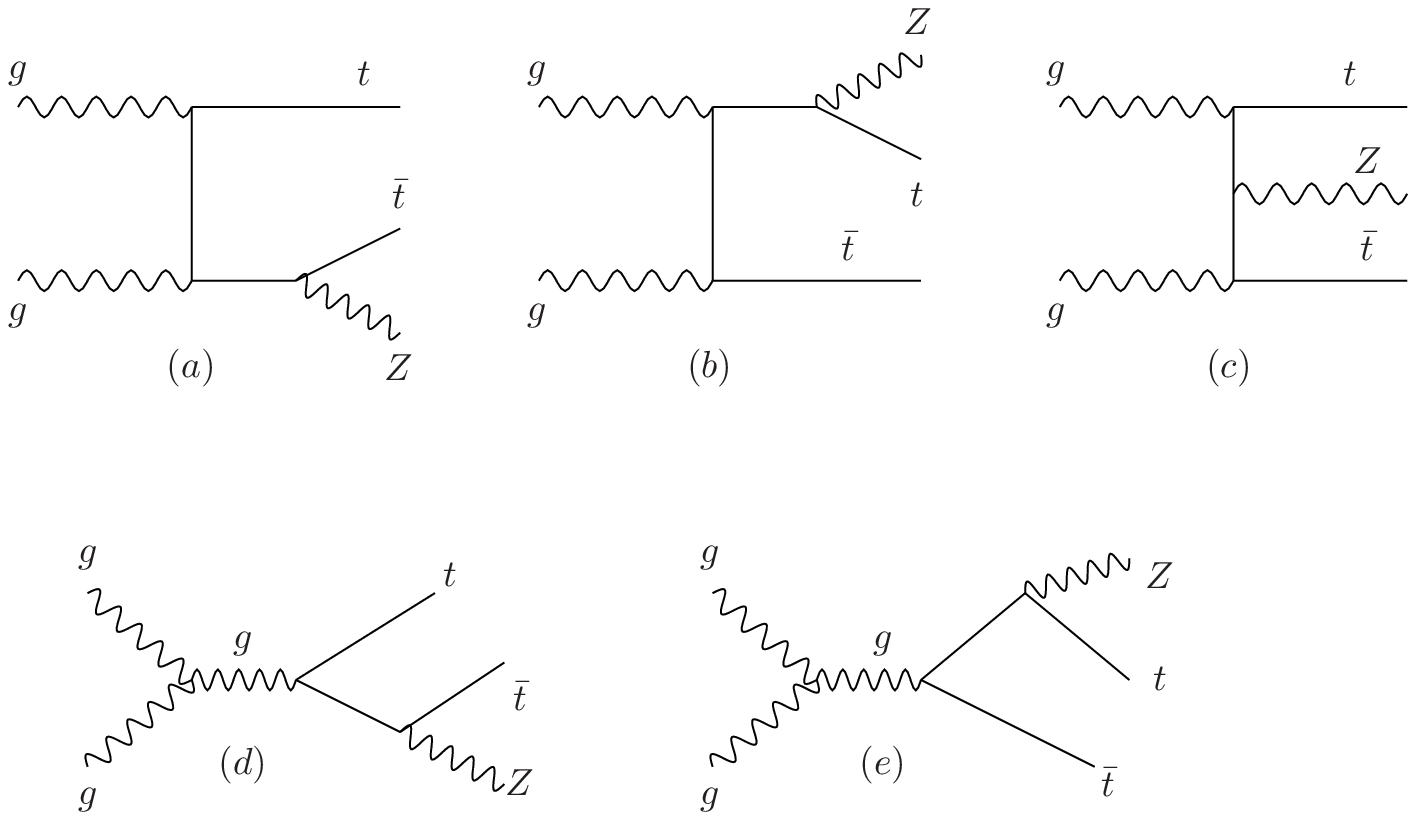 , height=9.cm}
\]\\
\vspace{-1.cm}
\hspace{4cm} SM diagrams for $gluon~gluon \to t\bar t Z$ process\\

\vspace{1cm}
\caption[1] {SM diagrams at Born level.}
\end{figure}

\clearpage

\begin{figure}[p]
\vspace{-1cm}
\[
\hspace{-2cm}\epsfig{file=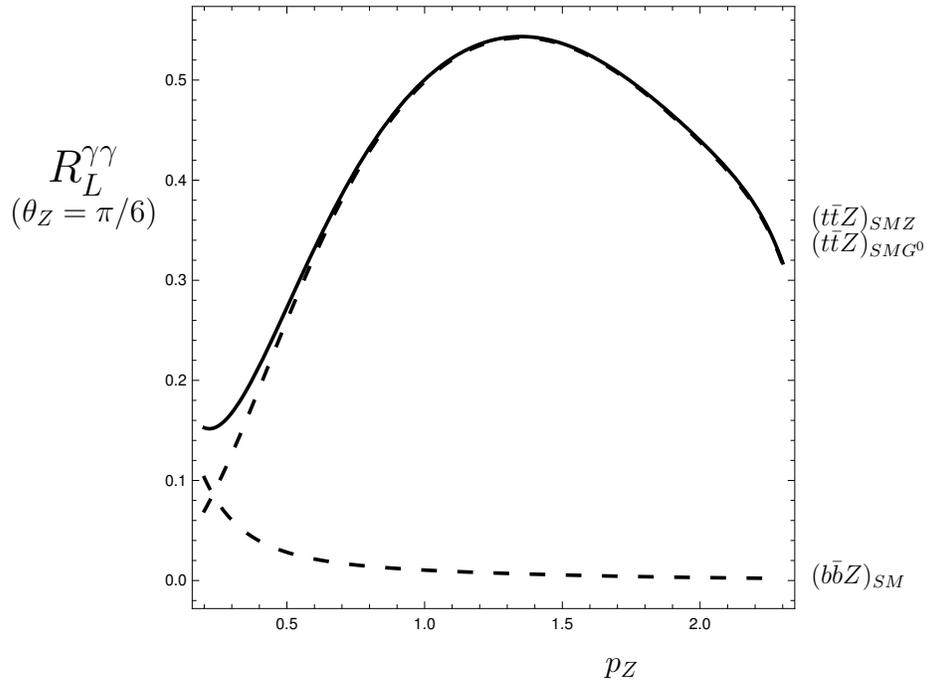 , height=9.cm}
\]\\
\vspace{0cm}
\[
\hspace{-2cm}\epsfig{file=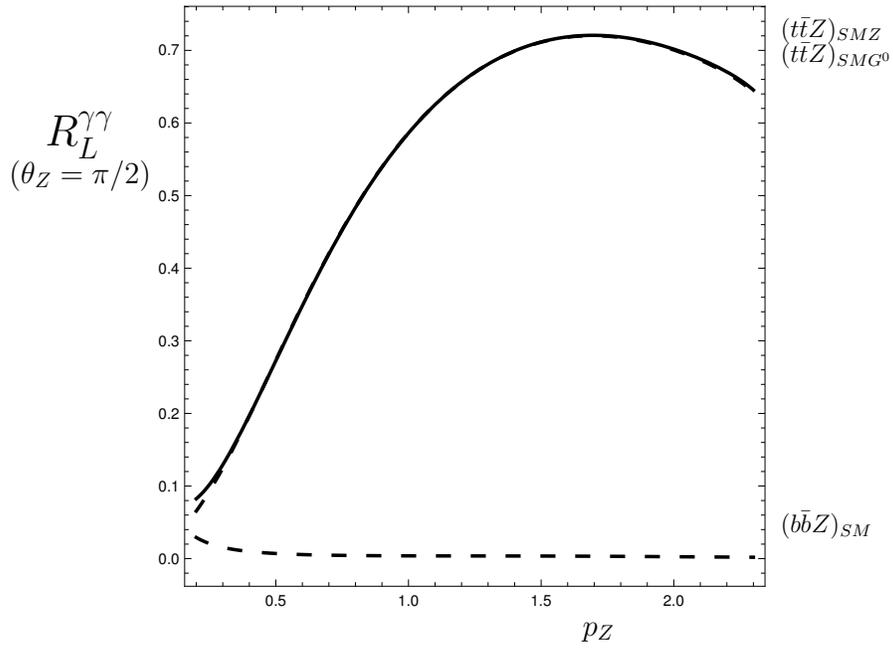 , height=8.5cm}
\]\\
\vspace{-1cm}
\caption[1] {SM $\gamma\gamma \to t\bar t Z_L$ ratio compared to the Goldstone case and to the $b\bar b Z_L$ one.}
\end{figure}

\clearpage

\begin{figure}[p]
\vspace{-2cm}
\[
\hspace{-2cm}\epsfig{file=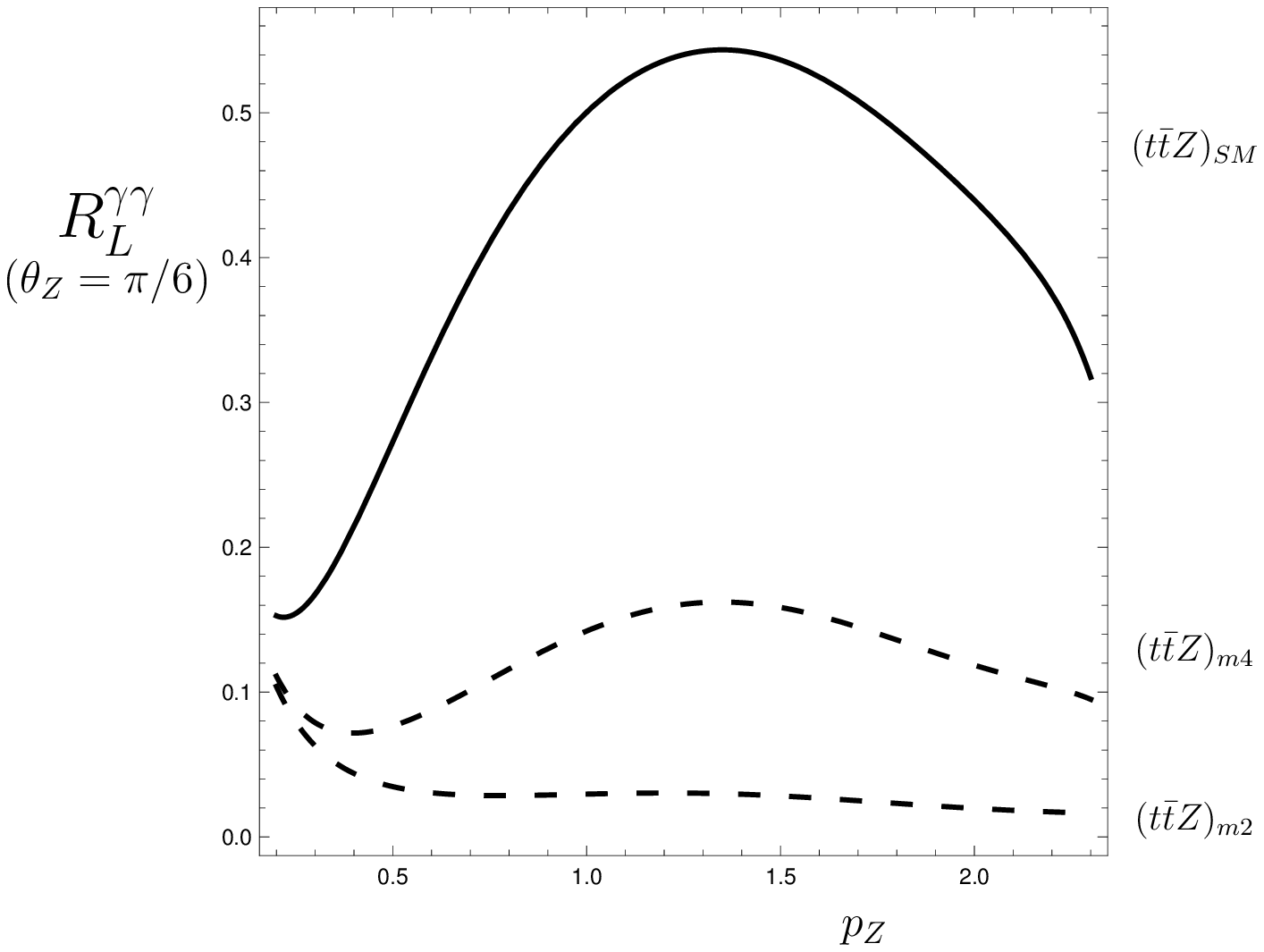 , height=9.cm}
\]\\
\vspace{0.5cm}
\[
\hspace{-2cm}\epsfig{file=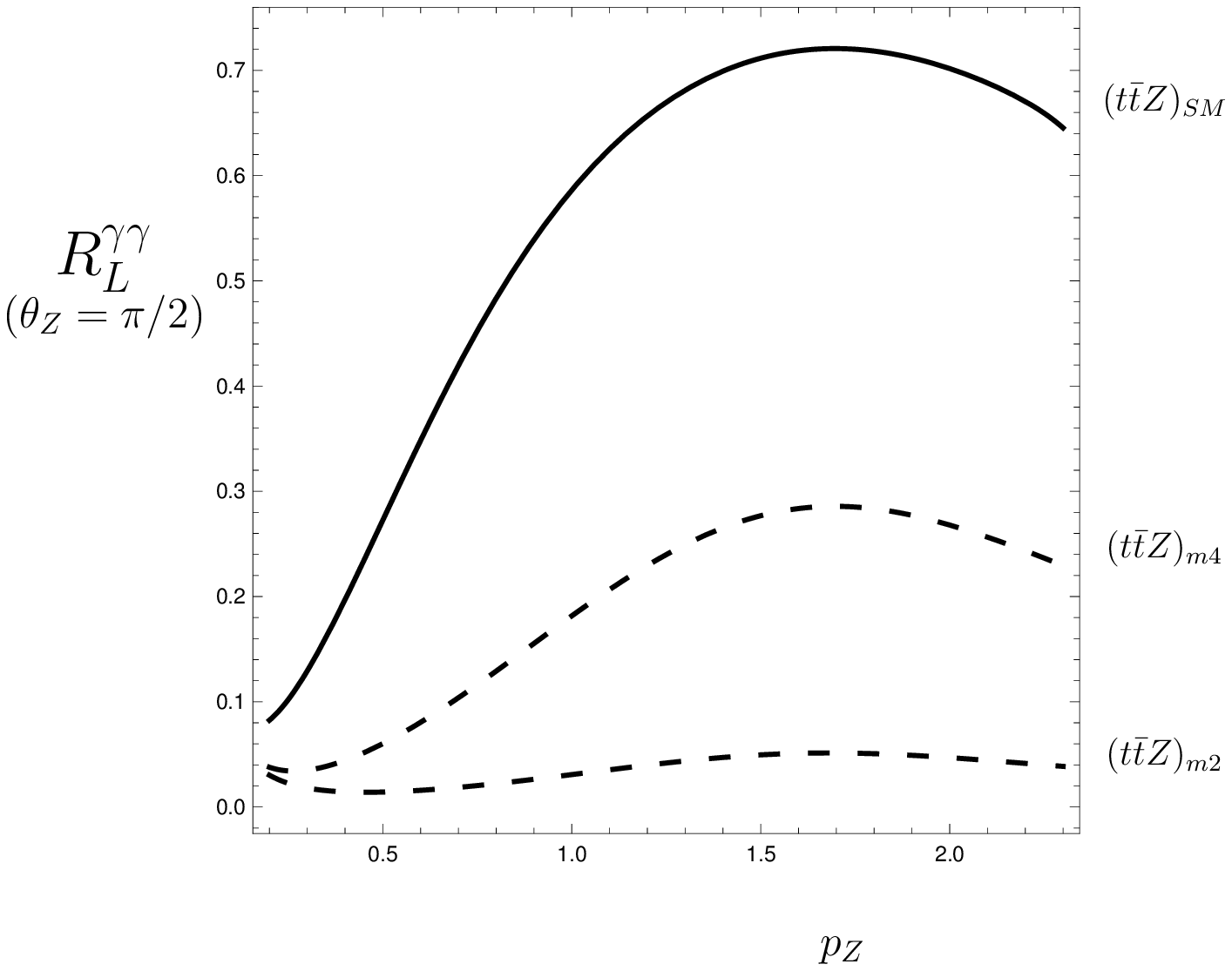 , height=9.cm}
\]\\
\vspace{-1cm}
\caption[1] {$\gamma\gamma \to t\bar t Z_L$ ratio for 2 cases of scale dependent top mass compared to the SM case.}
\end{figure}
\clearpage

\begin{figure}[p]
\vspace{-2cm}
\[
\hspace{-2cm}\epsfig{file=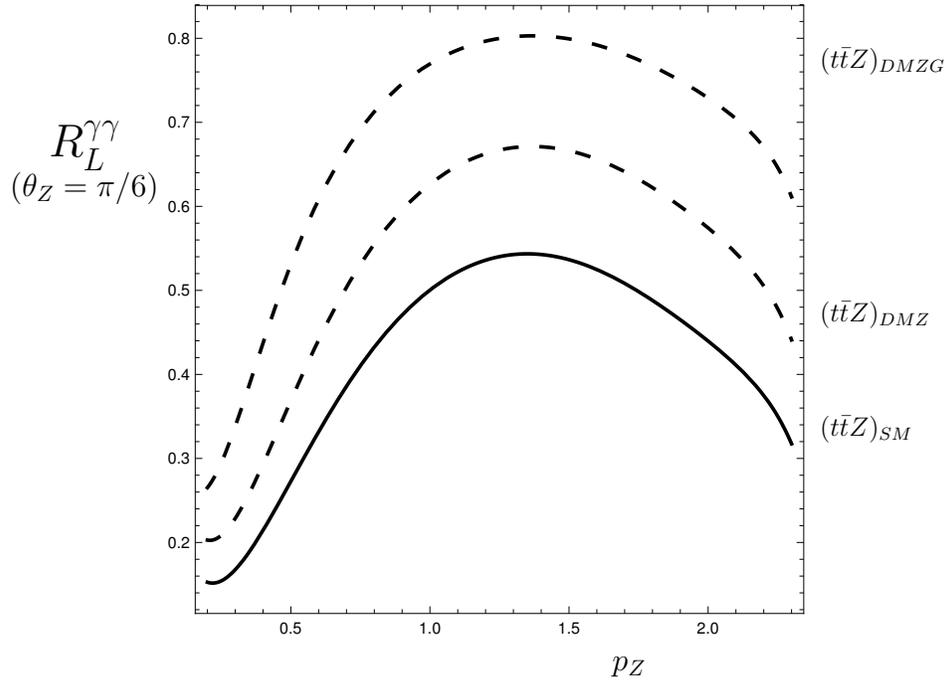 , height=9.cm}
\]\\
\vspace{1cm}
\[
\hspace{-2cm}\epsfig{file=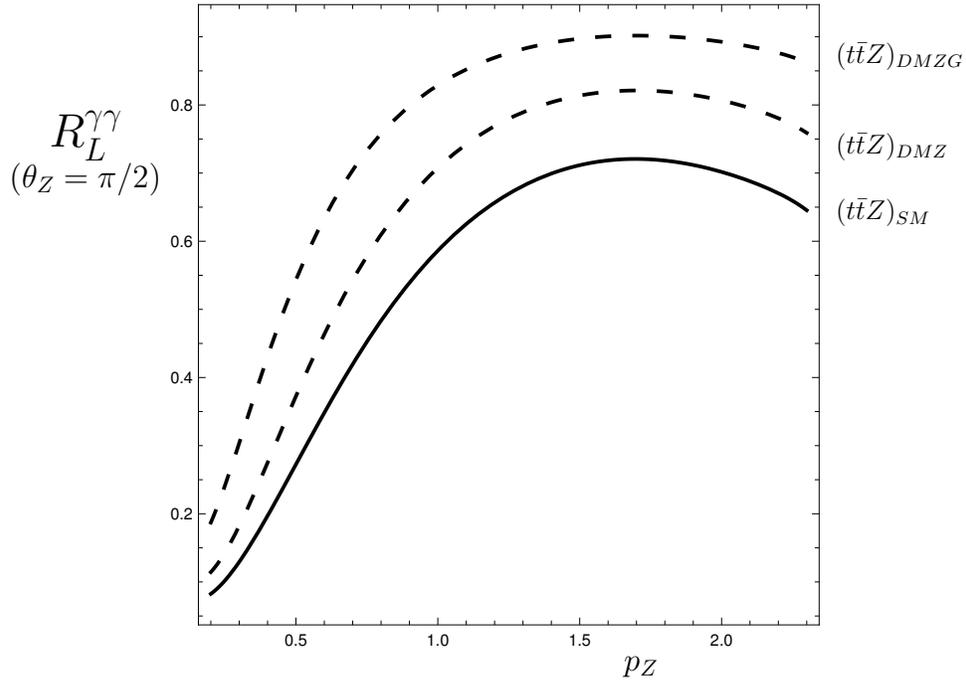 , height=9.cm}
\]\\
\vspace{-1cm}
\caption[1] {Photon-photon $Z_L$ ratio for 2 cases of Dark Matter final state interactions.}
\end{figure}
\clearpage

\clearpage

\begin{figure}[p]
\vspace{0cm}
\[
\hspace{-2cm}\epsfig{file=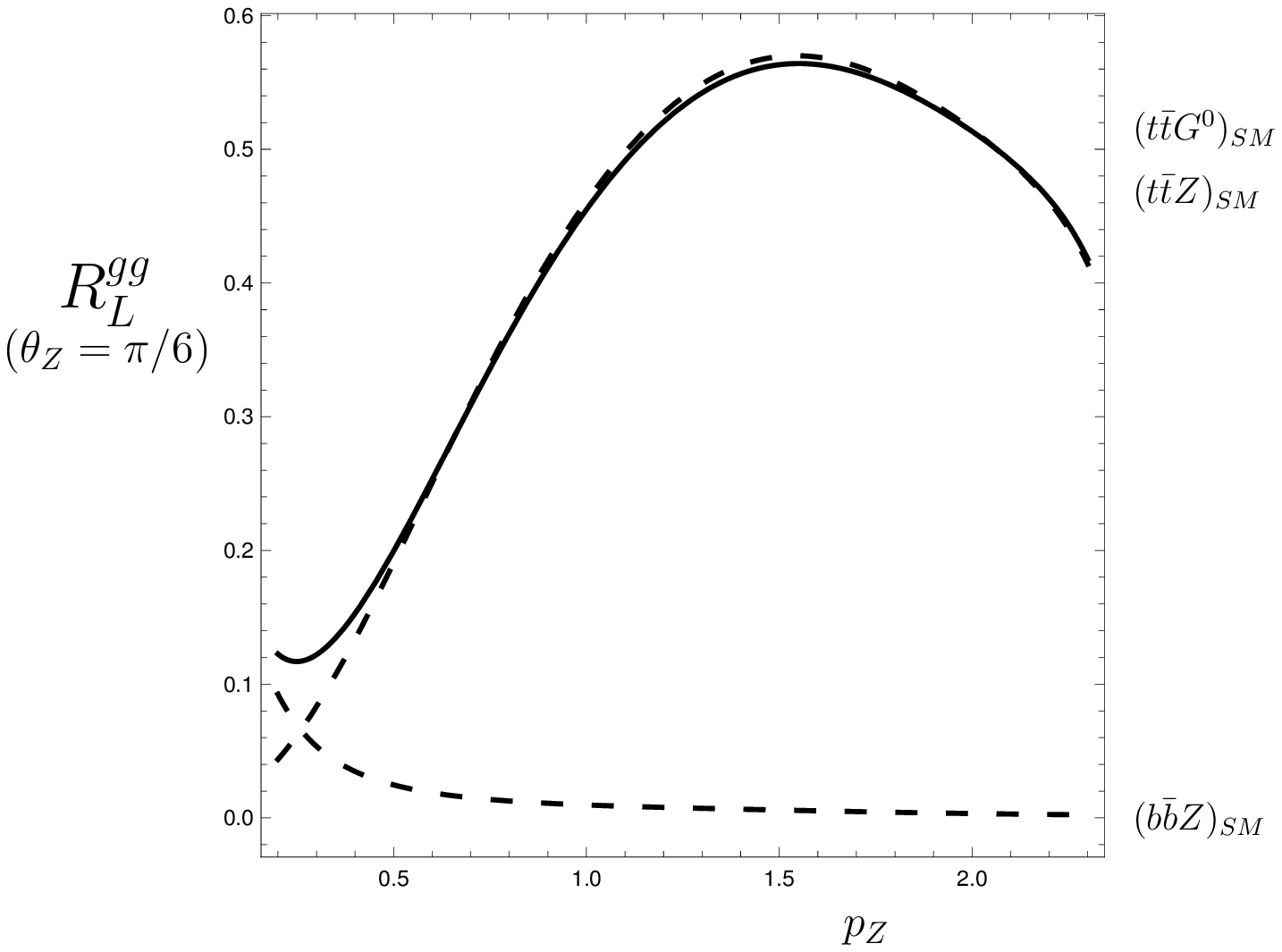 , height=9.cm}
\]\\
\vspace{0cm}
\[
\hspace{-2cm}\epsfig{file=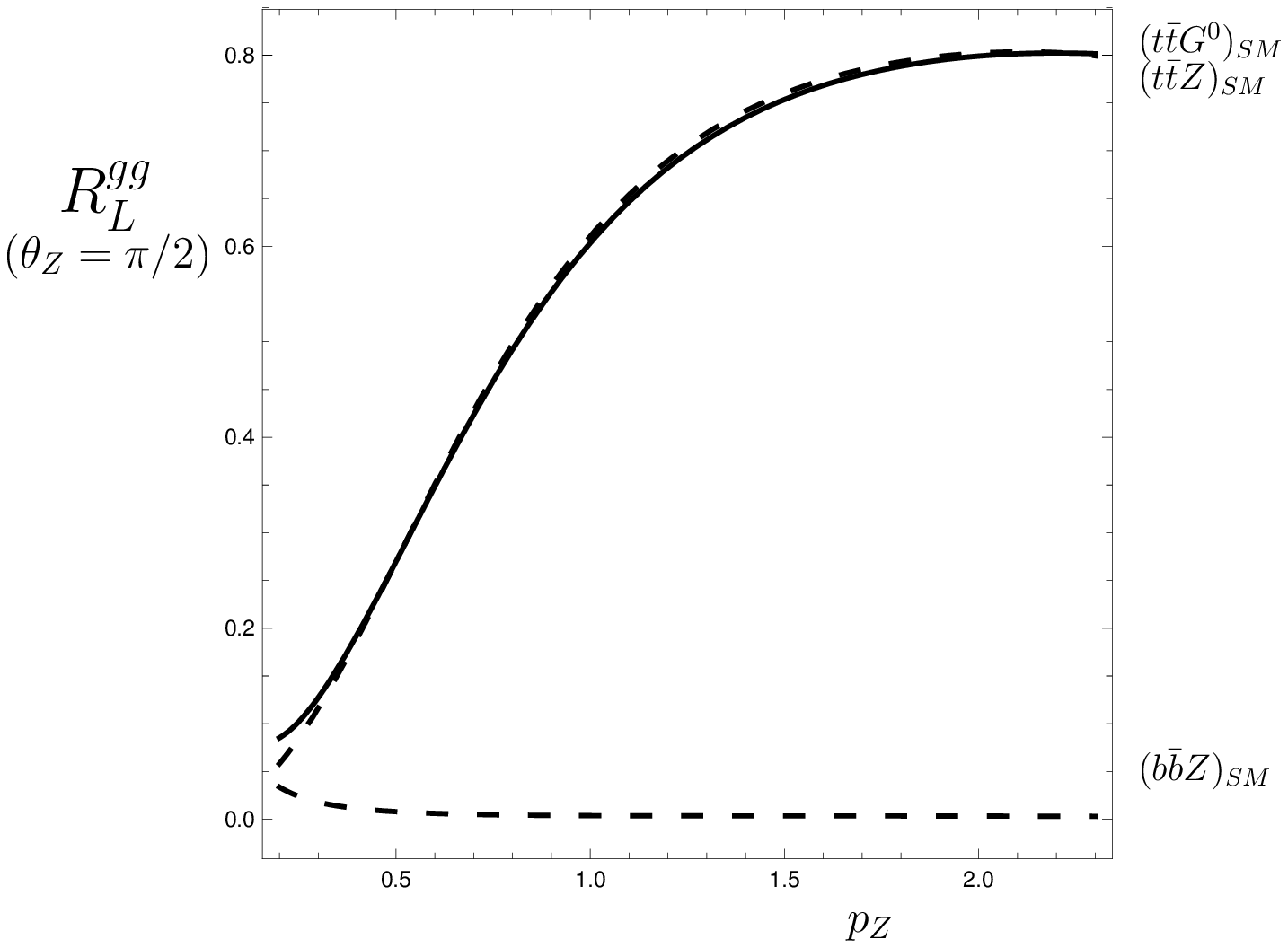 , height=9.cm}
\]\\
\vspace{-1cm}
\caption[1] {SM $gg\to t\bar t Z_L$ ratio compared to the Goldstone case and to the $b\bar b Z_L$ one  .}
\end{figure}

\clearpage

\begin{figure}[p]
\vspace{-1cm}
\[
\hspace{-2cm}\epsfig{file=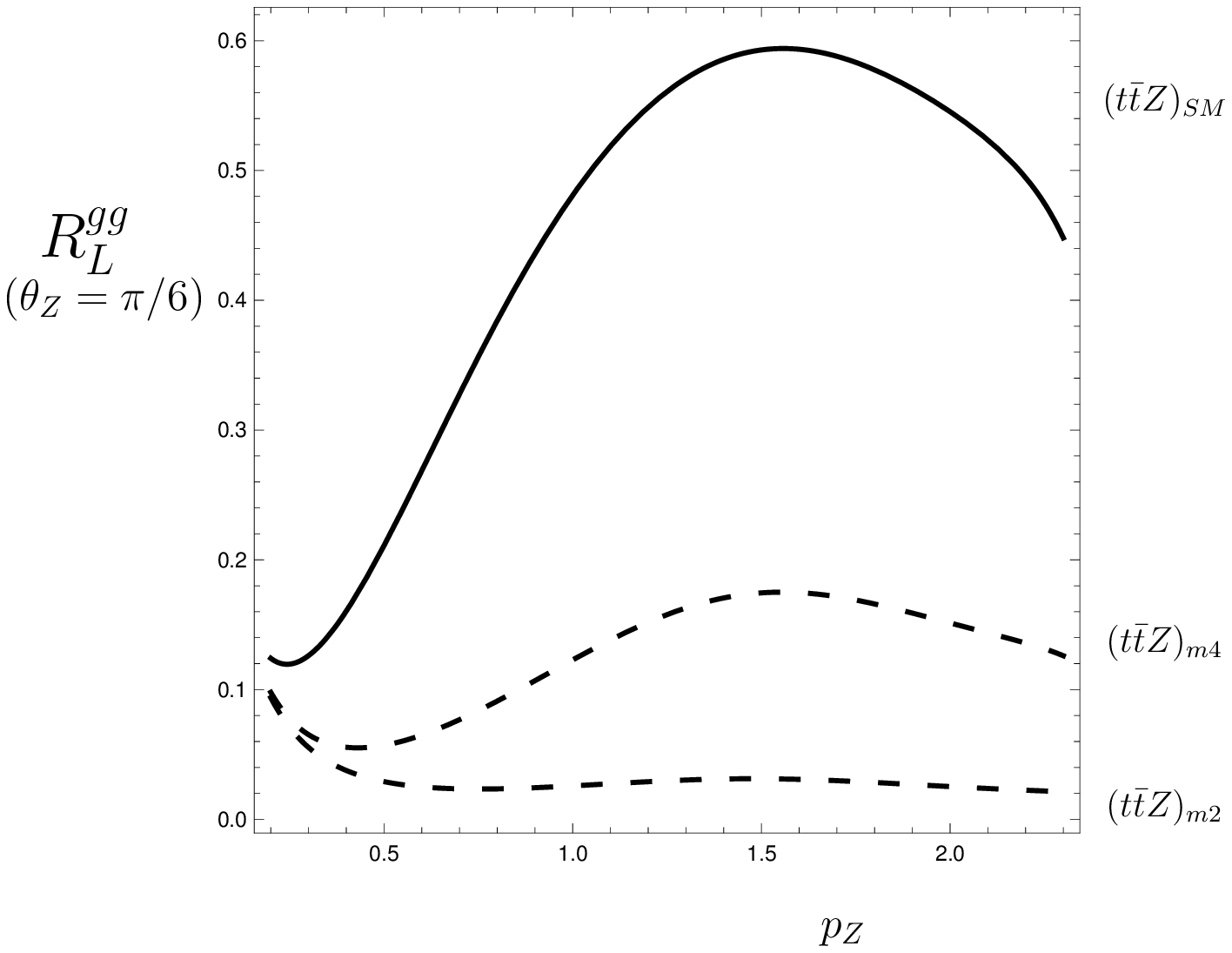 , height=9.cm}
\]\\
\vspace{0.5cm}
\[
\hspace{-2cm}\epsfig{file=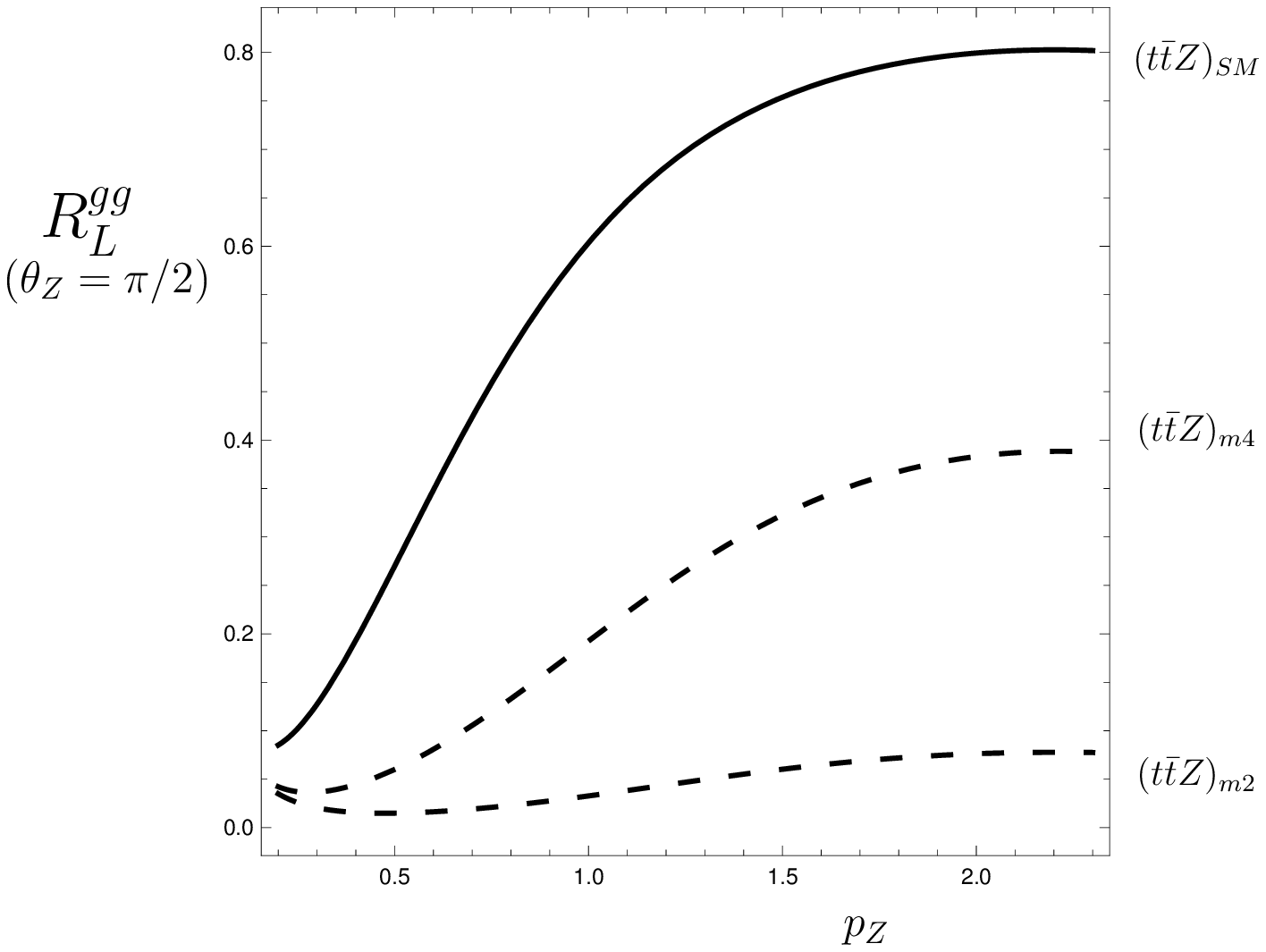 , height=9.cm}
\]\\
\vspace{-1cm}
\caption[1] {$gg \to t\bar t Z_L$ ratio for 2 cases of scale dependent top mass compared to the SM case.}
\end{figure}
\clearpage

\begin{figure}[p]
\vspace{-1cm}
\[
\hspace{-2cm}\epsfig{file=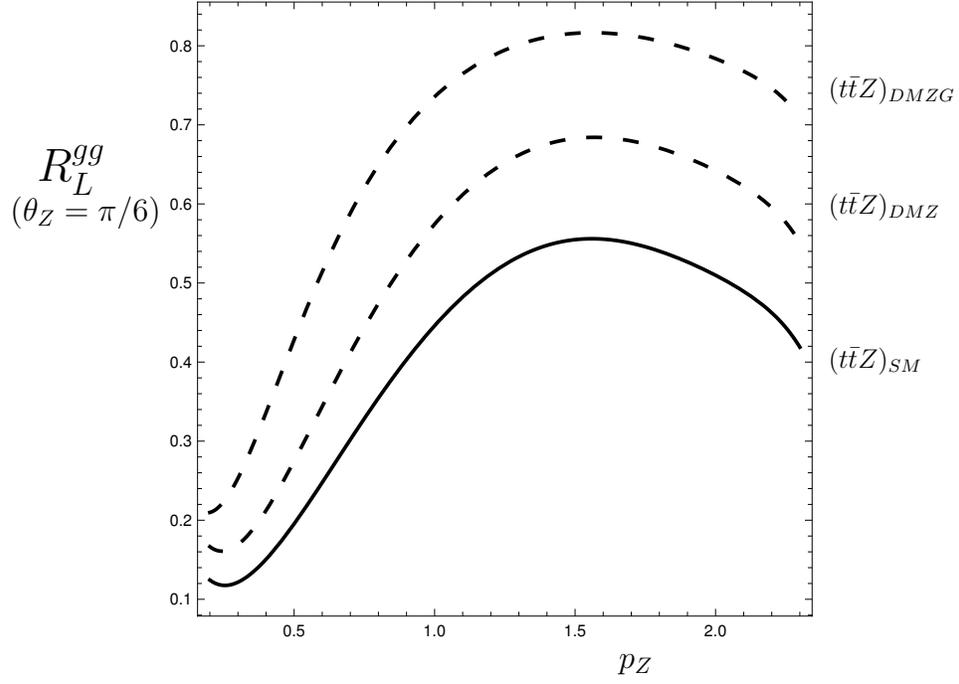 , height=9.cm}
\]\\
\vspace{0.5cm}
\[
\hspace{-2cm}\epsfig{file=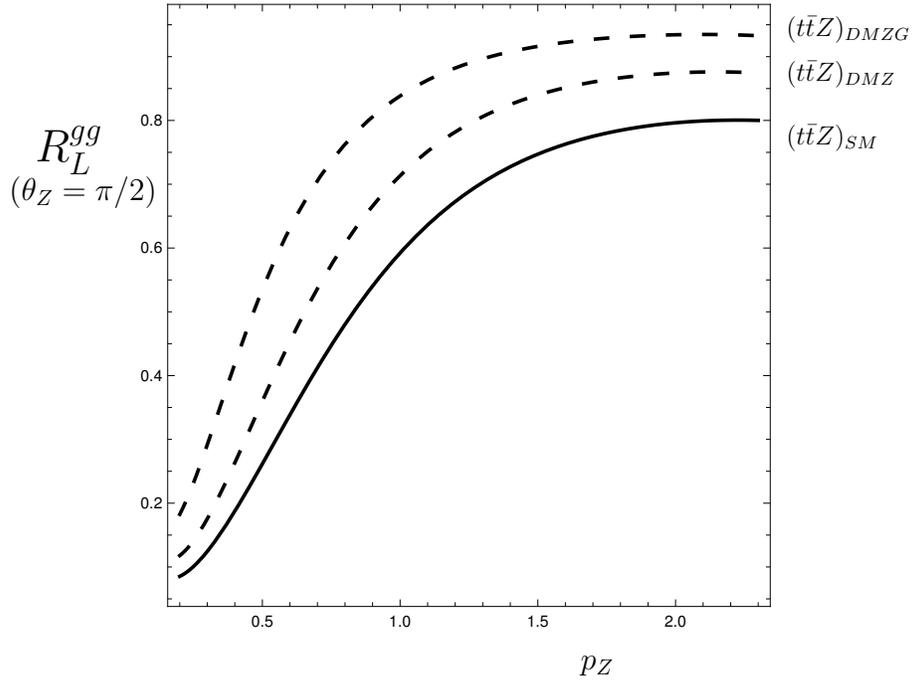 , height=9.cm}
\]\\
\vspace{-1cm}
\caption[1] {Gluon-gluon $Z_L$ ratio for 2 cases of Dark Matter final state interactions.}
\end{figure}
\clearpage

\clearpage

\end{document}